\documentclass[aps,prd,preprint,nofootinbib]{revtex4}
\usepackage{amsfonts}
\usepackage{mathrsfs}
\usepackage{graphicx}
\usepackage{amsmath}
\usepackage{amssymb}
\usepackage{epsfig}
\usepackage{graphicx}
\usepackage{slashed}
\usepackage{color}
\usepackage{multirow}
\usepackage{ulem}
\usepackage{adjustbox}
\newcommand{\mysslash}{\mathrel{\mkern-5mu\clipbox{0 0.75ex 0 0}{${\sslash}$}\!}}
\usepackage{stmaryrd}
\newcommand{\bqa}{\begin{eqnarray}}
\newcommand{\eqa}{\end{eqnarray}}
\newcommand{\beq}{\begin{equation}}
\newcommand{\eeq}{\end{equation}}
\allowdisplaybreaks[1]

\graphicspath{{fig/}{dia/}} \DeclareGraphicsExtensions{.eps}
\begin{document}

\title{Searching $B_c^\ast$ via conservation laws\\[0.7cm]}

\author{\vspace{1cm} Chia-Wei Liu \footnote[1]{chiaweiliu@ucas.ac.cn} and Bing-Dong Wan\footnote[2]{wanbingdong16@mails.ucas.ac.cn} \\}

\affiliation{School of Fundamental Physics and Mathematical Sciences, Hangzhou Institute for Advanced Study, UCAS, Hangzhou 310024, China\\
University of Chinese Academy of Sciences, 100190 Beijing, China
\vspace{0.6cm}} 

\begin{abstract}
\vspace{0.5cm}
To distinguish $B_c^{\ast +}(1^3S_1)$ and $B_c^+(1^1S_0)$ in the experiments, we propose two methods based on the conservation laws. 
\MakeUppercase{\romannumeral 1}. From the angular momentum conservation, a nonzero helicity of  $J/\psi$ of $B_c^{(\ast)+ } \to J/\psi \pi^+$ would be an evidence of $B_c^{\ast +}$. 
\MakeUppercase{\romannumeral 2}. Since $B_c^+ \to B^+ \phi$ is kinematically forbidden,  $ B_c^{\ast +} \to B^+ \phi$ provides a clean channel to probe $B_c^{\ast +}$\,.  
Particularly, our results show that $B_c^{\ast +}$ is promising to be observed at LHC  via $B_c^{(\ast)+} \to J/\psi \pi^+$. 
On the other hand, we find that ${\cal B} ( B_c^{\ast+} \to B^+ \phi)  = ( 7.0 \pm  3.0 ) \times 10 ^{-9} $, which is also feasible to be measured at the forthcoming experiments at HL-LHC and FCC-hh. 
\end{abstract}
\maketitle

\section{introduction}
The $B_c$ meson  is unique in the Standard Model (SM) as its members are composed of heavy quarks with two different flavors, beauty ($b$) and charm ($c$).
The $B_c$ mesons lie intermediate between ($c\bar{c}$) and ($b\bar{b}$) states both in mass and size, while the different quark flavors leads to much richer dynamics.
On the other hand, the ground state of $B_c$ mesons, unlike the charmonium and bottomonium, cannot annihilate into gluons or photons, providing an idea place to examine the heavy quarks.
Study on the $B_c$ mesons can deepen our understanding of both the strong and the weak interactions, revealing
the underlying physics of the heavy quark dynamics. Last but not least , it provides a unique hunting ground for searching new physics beyond the SM.

The ground state of $B_c$ meson was first observed by the CDF Collaboration at Fermilab~\cite{Abe:1998fb} in 1998, and there have been continuous measurements on both the  mass \cite{CDF:2007umr,LHCb:2012ihf,LHCb:2020ayi} and the lifetime \cite{CDF:2006kbk,D0:2008thm} via the exclusive decay $B_c^+\to J/\psi \pi^+$ and the semileptonic decay $B_c^+\to J/\psi l^+ \nu_l$. 
In 2014, the ATLAS Collaboration reported a structure with the mass of $( 6842\pm 9)$ MeV~\cite{Aad:2014laa}, which is consistent with the value predicted for $B_c(2S)$.
In 2019, the excited $B_c(2^1S_0)$ was confirmed and $B_c^\ast(2^3S_1)$ states have been observed in the $B_c^+\pi^+\pi^-$ invariant mass spectrum by the CMS and LHCb Collaborations, with their masses determined to be $(6872.1\pm 2.2)$  and $(6841.2\pm1.5)$ MeV \cite{Sirunyan:2019osb,Aaij:2019ldo}, respectively. 
The $B_c(2^1S_0)^+$ decays to $B_c^+(1^1S_0)\pi^+\pi^-$ directly, and the $B_c^\ast (2^3S_1)^+$ state decay to $B_c^{\ast +}(1^3S_1)\pi^+\pi^-$ followed by $B_c^{\ast +}(1^3S_1)\to B_c^+(1^1S_0)\gamma$. Since the soft photon in the intermediate decay $B_c^{\ast +}(1^3S_1)\to B_c^+(1^1S_0)\gamma$ was not reconstructed,  the mass of $B_c^\ast(2^3S_1)$ meson appears lower than that of $B_c(2^1S_0)$. 
This peculiar behaviors of the mass hierarchy makes $B_c^\ast(1^3S_1)$ uniquely important in studying the $B_c$ meson family.

\begin{figure}
	\centering
	\includegraphics[width=0.5\linewidth]{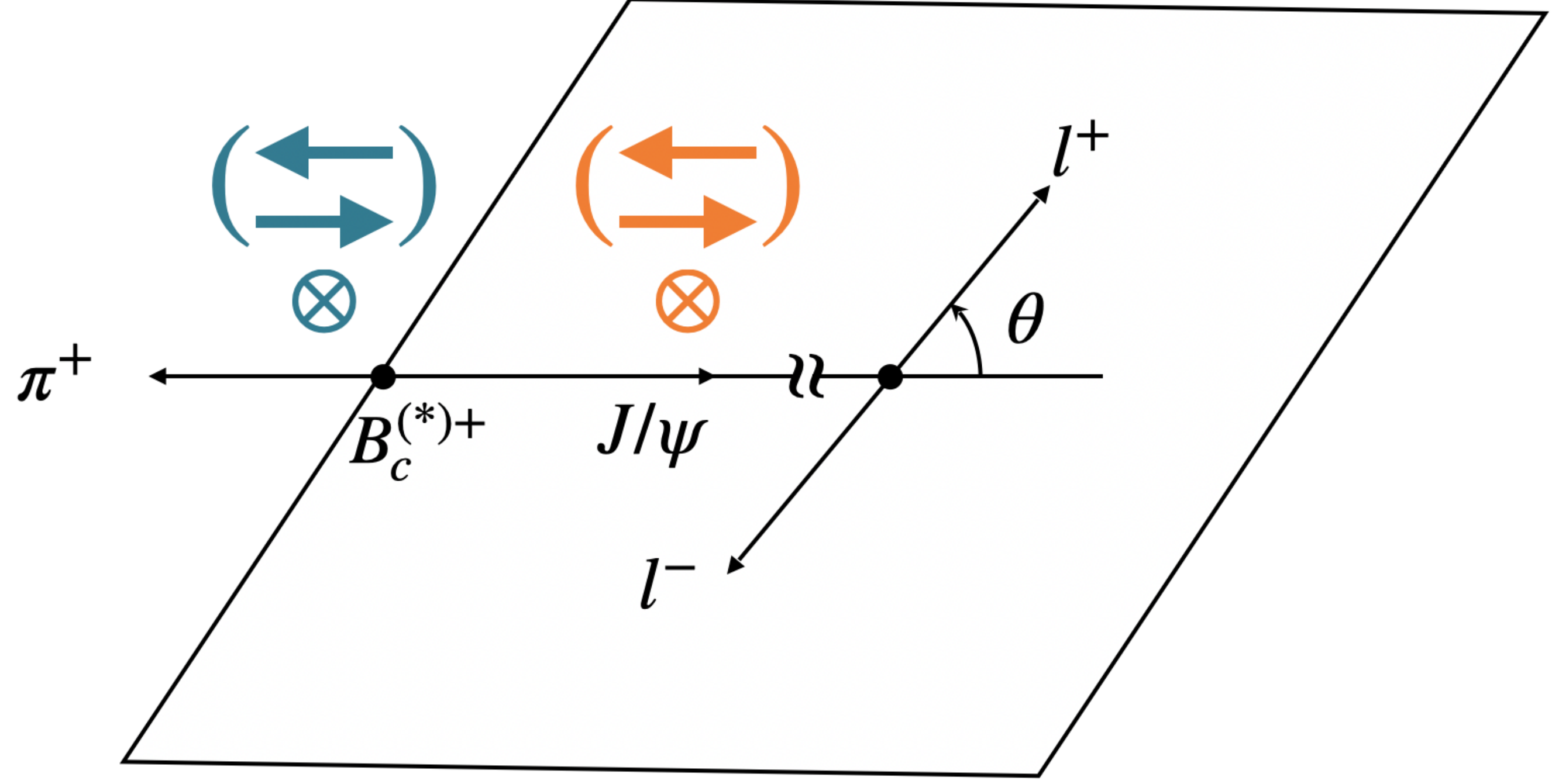}  
	\caption{The adjoint decay distributions of $B_c^{(\ast)+}  \to \pi^+J/\psi (\to l^+ l^-) $, where the blue and the orange represent the possible spin configuration(s) of $B_c^{(\ast)+}$   and $J/\psi$,  with  $\otimes$ indicating spin-0 at the  $\vec{p}_{J/\psi}$ direction. }
	\label{Deccay}
\end{figure}

\begin{figure}
		\includegraphics[width=.48\textwidth]{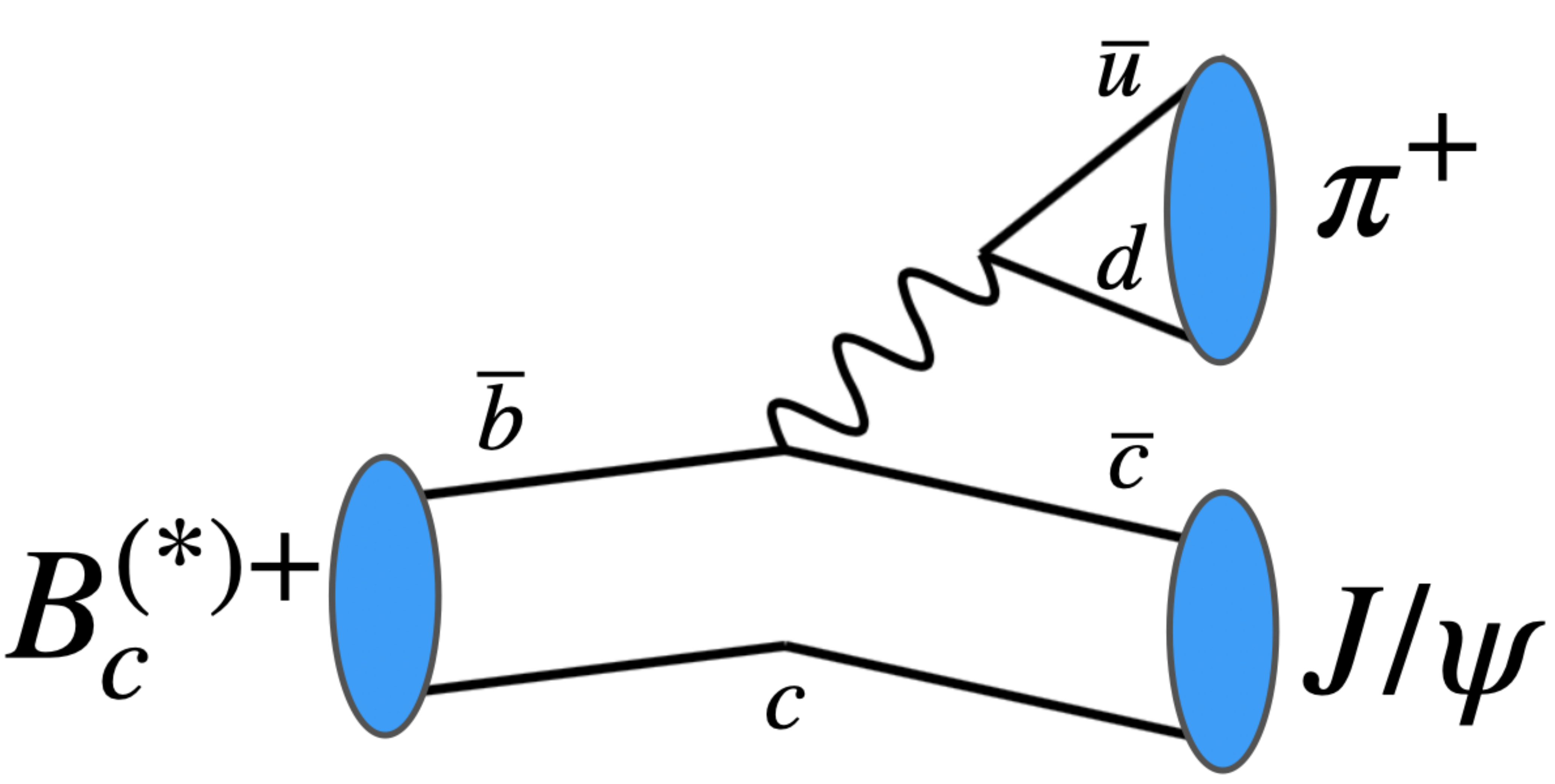}  
		\includegraphics[width=.48\textwidth]{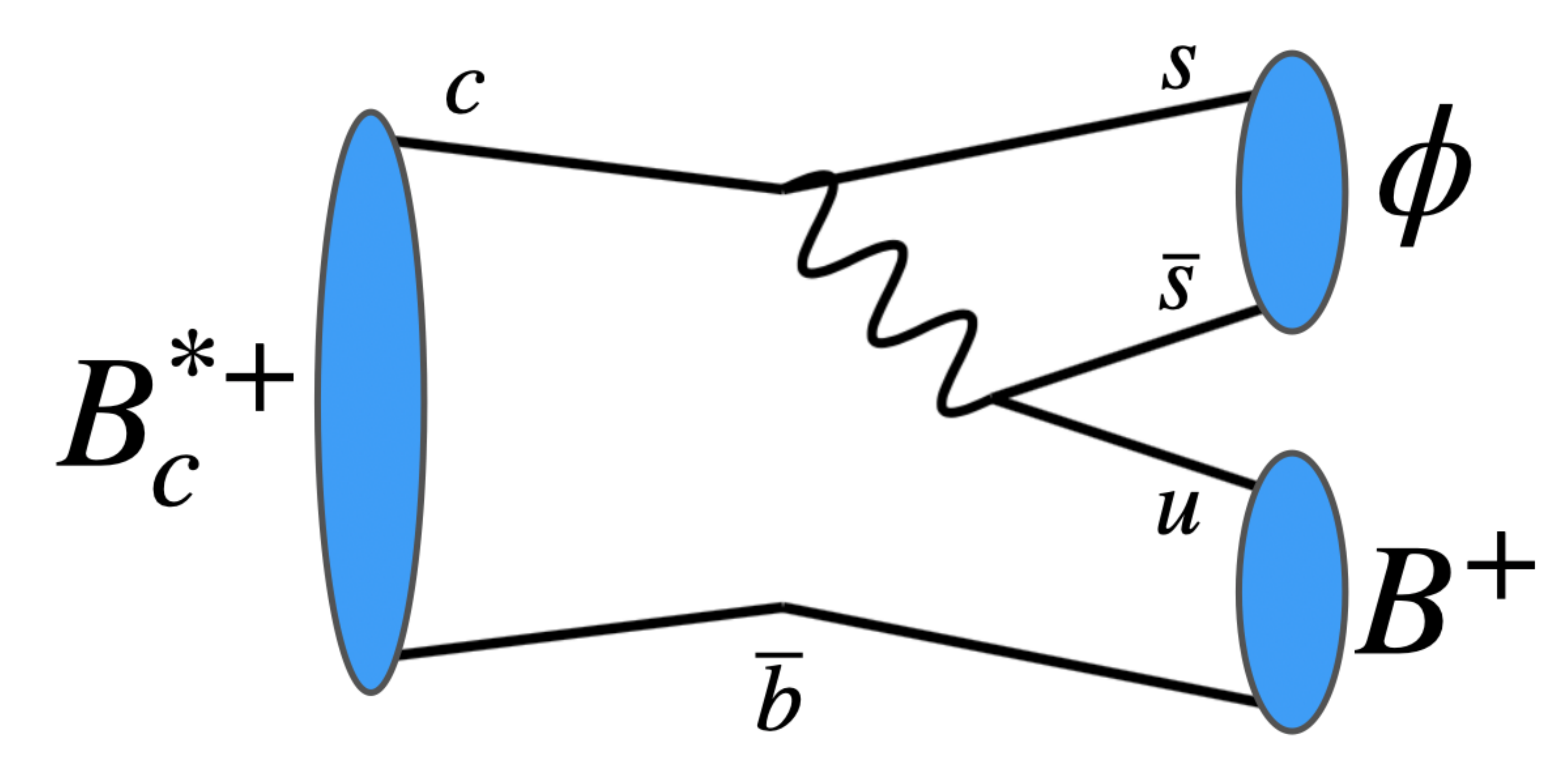}  
	\caption{  The quark diagrams for $B_c^{(*)+} \to J/\psi \pi^+$ and 
		$B_c^{+} \to B^+ \phi$ at the tree level.} 
	\label{Quark}
\end{figure}

In the following, we will abbreviate $B_c^\ast(1^3S_1)$ as $B_c^\ast$ so long as it does not cause confusion.
Study on the $B_c^\ast$ can complete the precise measurements of the spectrum of the  $B_c$ family, and the confirmation of its existence is of great importance for the understanding of strong interaction dynamics at low energy. 
On the mass of $B_c^\ast$, the theoretical predictions range discrepantly from $6326$ to $6346$ MeV \cite{Ding:2021dwh,Eichten:1994gt,Godfrey:2004ya,Mathur:2018epb,Li:2019tbn,Asghar:2019qjl,Eichten:2019gig}, and an experimental measurement is still lacking. 
The dominant decay mode $B_c^\ast\to B_c\gamma$ has not yet been observed, partly due to the noisy soft photon background of the hadron collider. 
To identify $B_c^{\ast}$ in the experiments, one of the important tasks is to distinguish them from $B_c$. 
In this study, we propose two methods based on the conservation laws:
\begin{itemize}
	\item  From the angular momentum conservation, the $J/\psi$ can only possess a zero helicity from $B_c^+\to J/\psi \pi^+$ as  $B_c^+$ is spin-0. In contrast, the $J/\psi$ of $B_c^{\ast +}\to J/\psi \pi^+$ can have either positive, zero, or negative helicities (see Fig.~\ref{Deccay}). 
	\item As $B_c^{+} \to B^+ \phi$ is kinematically forbidden, $B_c^{\ast +} \to B^+ \phi$ provides a clean channel. 
\end{itemize}
Their responsible quark diagrams at the tree level are given in Fig.~\ref{Quark}\,, where the hadronizations take place in the blue regions. As the W boson is color blind, the decays are color allowed and color suppressed, respectively.

The rest of the paper is organized as follows. The primary formulas in our calculation are presented in Sec. \ref{Formalism}. We give the numerical analysis and results in Sec. \ref{Numerical}. We conclude the study in Sec.~\ref{conclusion}.

\section{Helicity Formalism}\label{Formalism}

To extract the helicity information of $J/\psi$ as well as calculate the branching fractions,  we give the  helicity formalism of the decays in this section.
The helicity information of $J/\psi$ can be obtained  from  $B_c^{(*)+} \to J/\psi(\to l^- l ^+) \pi^+$ with $l = e, \mu$ . 
The advantage of 
the  helicity analysis is that 
it can easily cooperate with the sequential decays and has a clear view of physical meaning~\cite{Gutsche:2013oea}.

Taking the initial $B_c^{\ast +}$ as unpolarized, the angular distributions of $B_c^{(\ast)+} \to J/\psi (\to l^-l^+)\pi^+$ are given as 
\begin{equation}
\frac{\partial \Gamma^{(\ast)} }{\partial \cos \theta} \propto
\sum_{\lambda = \pm,0\,, l = \pm} \left| H_\lambda^{(\ast)}  d^1(\theta) ^{\lambda}\,_{l}\right|^2
\propto 1 - P_2 +\frac{3}{2} \alpha^{(\ast)}  P_2\,,
\end{equation}
where $H_\lambda^{(\ast)} $ are the helicity amplitudes with the subscripts denoting the helicity of $J/\psi$, $d^1(\theta)$  the Wigner $d$-matrix for $J=1$, 
$\theta$  defined in the helicity frame of $J/\psi$~(see Fig.~\ref{Deccay}), and
\begin{eqnarray}
&&P_2 = \frac{1}{2}\left(
3  \cos^2 \theta - 1 
\right)\,,\nonumber\\
&&\alpha^{(\ast)}  = \frac{|H_+^{(\ast)} |^2 + |H_-^{(\ast)} |^2 }{|H_+^{(\ast)} |^2 + |H_-^{(\ast)} |^2 + |H_0^{(\ast)} |^2 }\,.
\end{eqnarray}
Here, $\alpha$  has the physical meaning of the nonzero-polarized fraction of $J/\psi$. 
Notice that $H_\pm$ are forbidden by the angular momentum conservation, resulting in 
\begin{equation}\label{zero}
\alpha = 0 \,.
\end{equation}

To further extract the helicity information, we define 
\begin{equation} \label{add}
{\cal A} ^{(\ast)}  = \frac{1}{\Gamma} \left( 
\int _{|\cos \theta|<x_0} \frac{\partial \Gamma^{(\ast)}}{\partial \cos\theta} d\cos \theta- \int _{|\cos \theta|>x_0}\frac{\partial \Gamma^{(\ast)}}{\partial \cos\theta} d\cos \theta 
\right) = \left(3 x_0 - \frac{3}{2} \right) \alpha^{(\ast)} \,,
\end{equation}
where $x_0$ is chosen to satisfy 
\begin{equation}
x_0^3 - 3 x_0 +1 = 0 \,,
\end{equation}
 which is found to be $x_0\approx 0.3473$~.

The experiments of  $B_c^{\ast +}$ are polluted by the off-shell contributions from $B_c^+$ at LHC.
Thus, we define the event-average $\overline{ {\cal A}} $ as 
\begin{equation}\label{Abar}
\overline{ {\cal A} }  = r  {\cal A}  + r^* {\cal A}^{\ast}  =r^*{\cal A}^\ast \,,
\end{equation}
as well as the event-average nonzero-polarized fraction as
\begin{equation}\label{alphabar}
r \frac{\partial \Gamma }{\partial \cos \theta}  +  r^* \frac{\partial \Gamma^{\ast} }{\partial \cos \theta}  \propto 1 - P_2 + \frac{3}{2}\overline{\alpha} P_2 \,,
\end{equation}
with 
\begin{equation}
r^{(\ast)} = \frac{N_{B_c^{(\ast)}}}{N_{B_c} + N_{B_c^{\ast}}}\,,
\end{equation}
where $N_{B_c^{(\ast)} } $ is the number of the observed events in $B_{c}^{(\ast)+} \to J/\psi(\to l^+ l^- )  \pi^+$.
The second equality in Eq.~\eqref{Abar} is  attributed to Eq.~\eqref{zero}\,.  

To get an estimation on the experiments, we calculate the amplitudes within the factorization framework.
The helicity amplitudes of $B_c^{(\ast)+} \to J/\psi \pi^+$ are given as 
\begin{eqnarray}
 &&(2\pi )^4 \delta^4(p_{B_c} - p_{J/\psi} -p_{\pi}) H_\lambda =\nonumber\\ &&~~~~i\frac{G_F}{\sqrt{2}}V_{cb}^\ast V_{ud}  f_\pi p_\pi^\mu a_1 \langle J/\psi; p\hat{z}, J_z = \lambda| \overline{b}\gamma_\mu ( 1 - \gamma_5) c|B_c^{(\ast)+}; J_z = \lambda \rangle \,,
\end{eqnarray}
where $p$ is for the 4-momentum of the hadron in the subscript, $G_F$  and $f_\pi$  the Fermi and the pion decay constants,  $a_1$  the effective Wilson coefficient for the color-allowed decays, $J_z$  the angular momentum at the $z$ direction, and $p\hat{z}$ indicates $\vec{p}_{J/\psi} \mysslash \hat{z}$. 

On the other hand, the helicity amplitudes of  $B_c^{\ast +} \to B^+ \phi$ are given as  
\begin{eqnarray}
&&(2\pi )^4 \delta^4(p_{B_c} - p_{B^+} -p_{\phi}) H_\lambda =\nonumber\\ 
 &&~~~~ - \frac{G_F}{\sqrt{2}}V_{cs}^\ast V_{su}  f_\phi \epsilon^{\mu\ast}_{\lambda} a_2 \langle B^+; p\hat{z}| \overline{u}\gamma_\mu ( 1 - \gamma_5) c|B_c^{(\ast)+}; J_z = -\lambda \rangle \,,
\end{eqnarray}
where $ a_2$ is the effective Wilson coefficient for the color-suppressed decays,  $f_\phi $  the  $\phi$ decay constant, and $\epsilon^{\mu \ast}_{\lambda}$   the polarization 4-vector of $\phi$ with $\lambda$ its helicity. 

Finally, the decay width for $B_c^+$ is given as 
\begin{equation}
\Gamma = \frac{|\vec{p}_{\text{cm}}| }{8 \pi M_{B_c}} \left| H_0\right|^2   \,,
\end{equation}
whereas the decay widths  of $B_c^{*+}$ with the  daughter vector meson  having $\lambda$ helicity are given as 
\begin{equation}
\Gamma_\lambda  = \frac{|\vec{p}_{\text{cm}}| }{24 \pi M_{B_c^\ast}} \left| H_\lambda^\ast \right| ^2  \,.
\end{equation}
The total decays widths of $B_c^\ast \to J/\psi \pi^+$ and $B_c^\ast \to B^+ \phi $ can be easily obtained by adding up the contributions from $\lambda = 0 , \pm$ .

\section{Numerical analysis}\label{Numerical}

The meson transition matrix elements  require the knowledge of the hadron wave functions. In this work,  we employ the ones from  the homogeneous bag model,
in which the center motions of the hadrons in the original bag model are removed~\cite{Geng:2020ofy}. 
The bag radius~$(R)$ and the quark masses can be extracted from the mass spectra, which are found to be~\cite{Zhang:2021yul}
\begin{equation}
R = ( 2.81 \pm 0.30) ~\text{GeV}^{-1}\,,~~~M_{u,d} =0\,,~~~M_c = 1.641~\text{GeV}\,,~~~M_b = 5.093~\text{GeV}\,.
\end{equation}
The details of the calculation can be found in the Appendix. 
In this study, $f_\pi$ and $f_\phi$ are taken from the experiments and  the Lattice QCD~\cite{Chen:2020qma,pdg} 
\begin{equation}
f_\pi = 131~\text{MeV}\,,~~~f_\phi  = (241\pm 9 )~\text{MeV}\,,
\end{equation}
and the effective Wilson coefficients are taken to be
\begin{equation}
|a_1| = 1.0 \pm 0.1\,,~~~~|a_2| = 0.27 \pm 0.07\,.
\end{equation}
The results  are given in Table~\MakeUppercase{\romannumeral 1}\,, where we also include $\Gamma( B_c^{\ast +} \to B_c^+ \gamma )$, which can be safely approximated as $1/\tau$ with $\tau$ the lifetime of $B_c^{\ast +}$.
The calculated lifetime is consistent with most of the literature~\cite{Li:2019tbn,consistent}, but significantly smaller than  the one from the nonrelativistic potential model~\cite{Eichten:1994gt}, and twice larger than the one from the relativistic independent quark model~\cite{twice}.  
Nonetheless, a large part of the uncertainties that arises from the hadron wave functions is canceled in the branching ratios of $B_c^{\ast +}$, as the lifetime is calculated under the same framework.

\begin{table} 
	\caption{The decay widths and the branching ratios.	}
\label{table}
	\begin{tabular}[t]{lcccc}
		\hline
		\hline
		Channel &Helicity & $\Gamma ~ ( \text{eV})$  & ${\cal B}$\\
		\hline
		$ B_c^+ \to J/\psi \pi^+ $&$H_0$& $ (6.3\pm 1.3)\times 10^{-7} $&$(4.8 \pm 1.0)   \times 10^{-4}$\\
		\hline
 \multirow{4}{*}{
$ B_c^{\ast+} \to J/\psi \pi^+   $ }& $H_-$&$ (6.4\pm 2.7)\times 10^{-9}$&$(1.2\pm 0.5)  \times 10^{-10}$\\
& $H_0$ ~& $( 2.8 \pm 0.6)\times 10^{-7}$&$(5.5\pm1.4) \times 10^{-9}$\\
& $H_+$& $ (9.9 \pm 2.3 ) \times 10^{-7}$&$(1.9 \pm 0.5) \times 10^{-8}$\\
 &Total& $(1.2  \pm 0.2 ) \times 10^{-6}$ &$ (2.4 \pm 0.5)  \times 10^{-8}$ \\
		\hline
$B_c^{\ast +} \to B_c^+ \gamma $&Total&$  53 \pm 3 $&$\approx$ 1 \\
$B_c^{\ast +} \to B^+ \phi $&Total&$(3.7\pm 1.7 )\times 10^{-7}$&$(7.0 \pm 3.0)\times 10 ^{-9} $  \\
\hline
\hline
	\end{tabular}
\end{table}

Our  ${\cal B} (B_c^+ \to J/\psi \pi^+)$ is consistent with  the relativistic quark model~\cite{BcRQ}, but two times smaller compared to most of the literature~\cite{Jpsipip}\,, which can be partly attributed to that we use a smaller $|a_1|$. 
 As our estimation is a more conservative one,  the angular analysis is promising to be carried out in the experiments for there are more data points to reconstruct the distribution than we expect.

\begin{figure}
	\centering
	\includegraphics[width=0.5\linewidth]{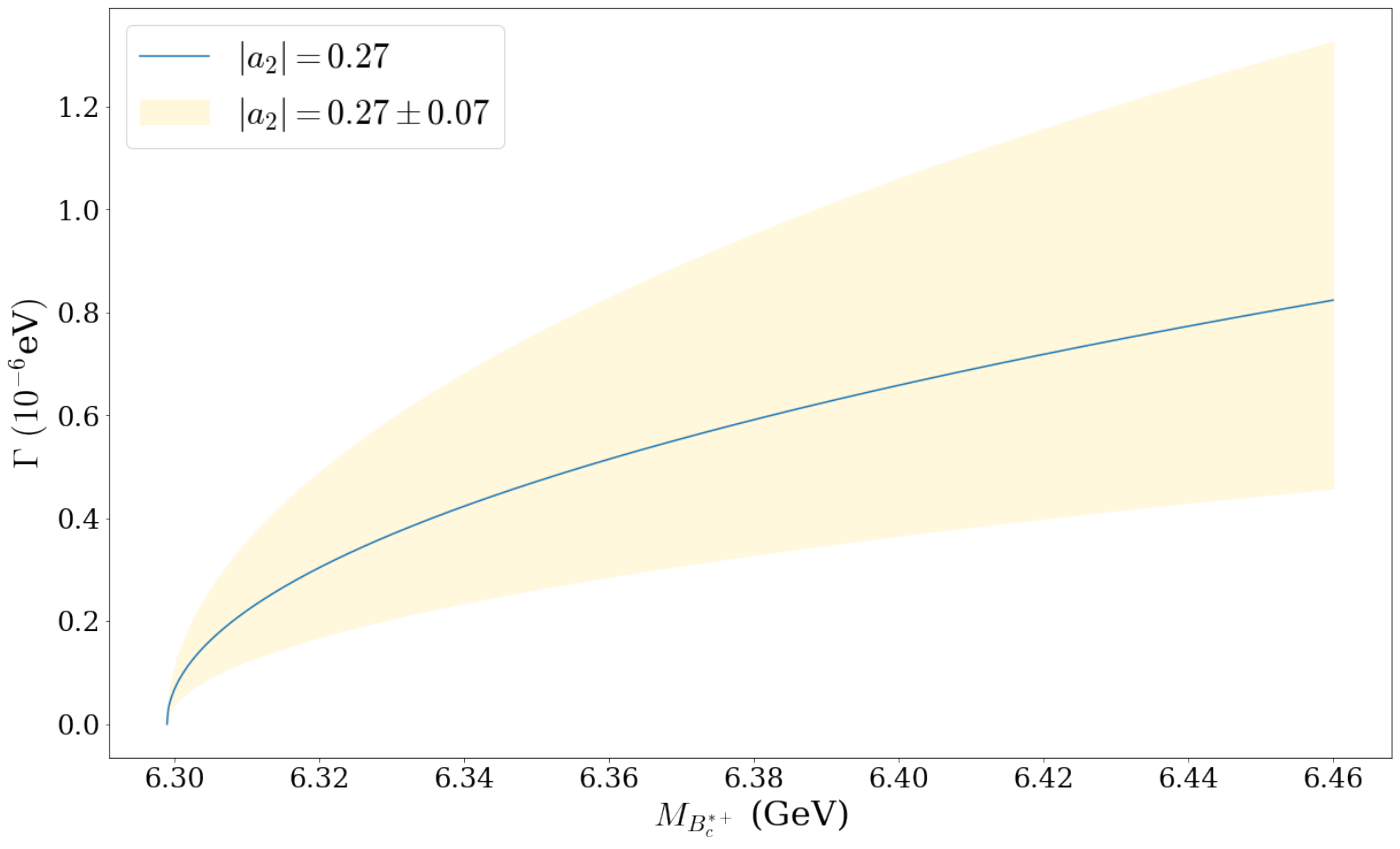}  
	\caption{ $\Gamma( B_c^{\ast +} \to B^+ \phi) $ versus $M_{B_c^{\ast +}} $, where the yellow region covers the uncertainty of $a_2$.}
	\label{a2}
\end{figure}

The decay of $B_c^{\ast +} \to B^+ \phi$ is color suppressed and suffers large uncertainties from $a_2$ as well as $M_{B_c^\ast}$.
In particular, as $M_{B_c^{\ast }}$ is close to the mass threshold of $B^+ \phi$, the decay width can range from $0$ to $10^{-6}$~eV, depending on $M_{B_c^{\ast +}}$.  The  dependency on  $M_{B_c^{*+}}$ as well as the uncertainties caused by $a_2$ are plotted in Fig.~\ref{a2}.
Taking $M_{B_c^{\ast +} }= 6331$~MeV, the calculated decay width is given in Table~\ref{table}, which is consistent with Ref.~\cite{Search}\,, within the range of the error. 

From  Table~\ref{table}, for $B_c^{\ast+} \to J/\psi \pi^+$ we obtain 
\begin{equation}
\alpha^\ast =  0.82 \pm 0.01 \,,~~~{\cal A}^\ast = 0.38 \pm 0.01\,,
\end{equation}
in which  the  theoretical uncertainty is canceled for the correlations between $H_\lambda^\ast$.  The cross section of $B_c^\ast$ meson at the LHC is expected to be $\sigma\left({B_c^\ast}\right)=29\ {\rm nb}$ \cite{Chang:1992jb}. At an integrated luminosity of $150\ \text{fb}^{-1}$ during LHC Run-2, $300\ \text{fb}^{-1}$ during LHC Run-3, and $3000\ \text{fb}^{-1}$ after High Luminosity upgrade (HL-LHC)~\cite{Apollinari:2017lan}, the numbers of $B_c^\ast$ events are $8.7\times 10^9$, $1.74\times 10^{10}$ and $1.74\times 10^{11}$, resulting in $270$, $540$, and $5400$ events of  $B_c^{\ast +}\to J/\psi \pi^+$, respectively. 
Taking the branching ratios ${\cal B} (J/\psi \to l^+ l ^ - )\approx 12\%$~\cite{pdg}, there are expected to be
 $33$, $65$, and $650$ events of $B_c^{\ast +}  \to \pi^+J/\psi (\to l^+ l^-) $  being able to be  reconstructed at LHC Run-2, LHC Run-3 and HL-LHC, respectively. 

By choosing 
$6400~\text{MeV}> M(J/\psi \pi^+) > 6325~\text{MeV}$
with $M(J/\psi \pi^+)$ the invariant mass of $J/\psi \pi^+$, most of the off-shell contribution from $B_c^+$ would be filtered, in which $N_{B_c}$ is expected to be less than 20 at running LHC from the  Fig.~1 of Ref.~\cite{LHCb:2016vni}\footnote[1]{In fact, at the bottom right figure, there appears to have a little bump around $6340$ MeV.}. 
Thus, $N_{B_c}$ and $N_{B_c^*}$ can be safely taken as equal in the simulation.

\begin{figure}
 	\includegraphics[width=0.45 \textwidth ]{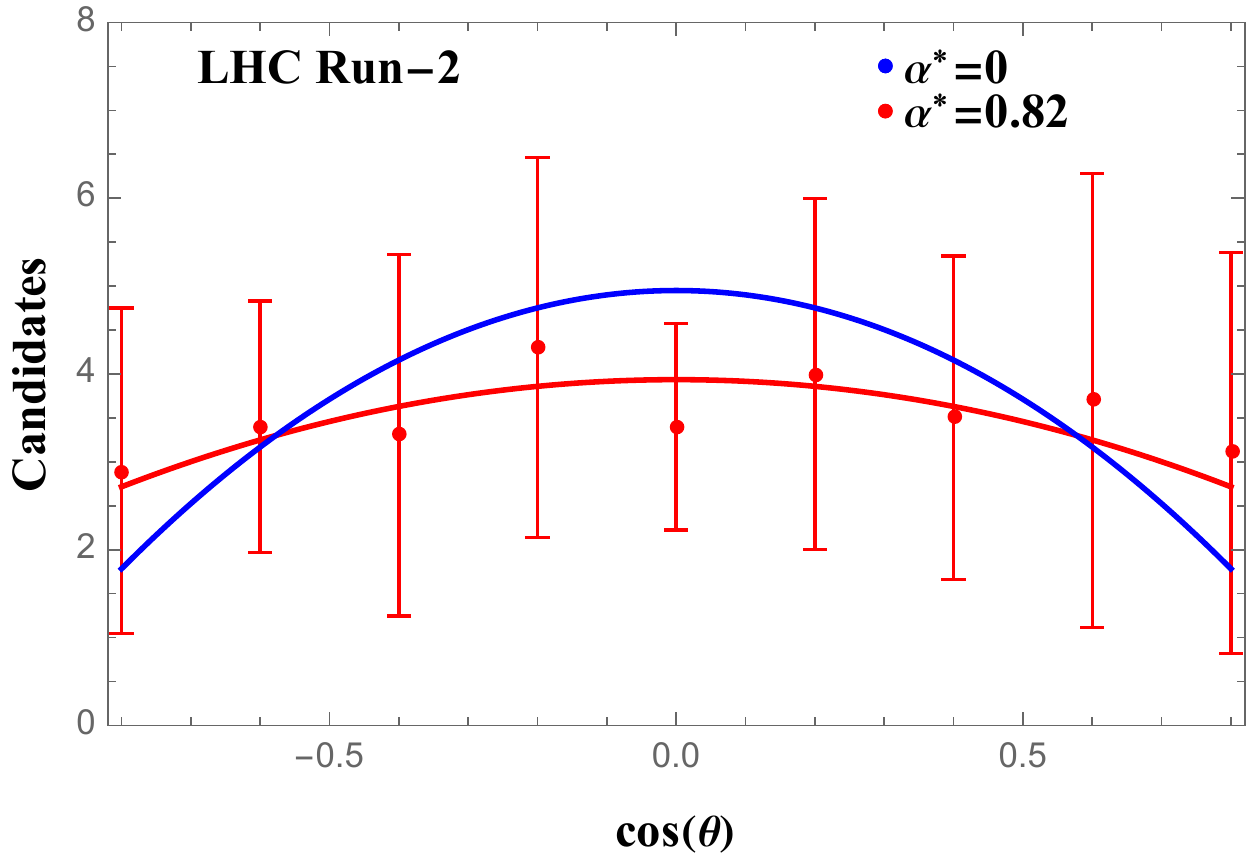}
 	\includegraphics[width=0.45 \textwidth ]{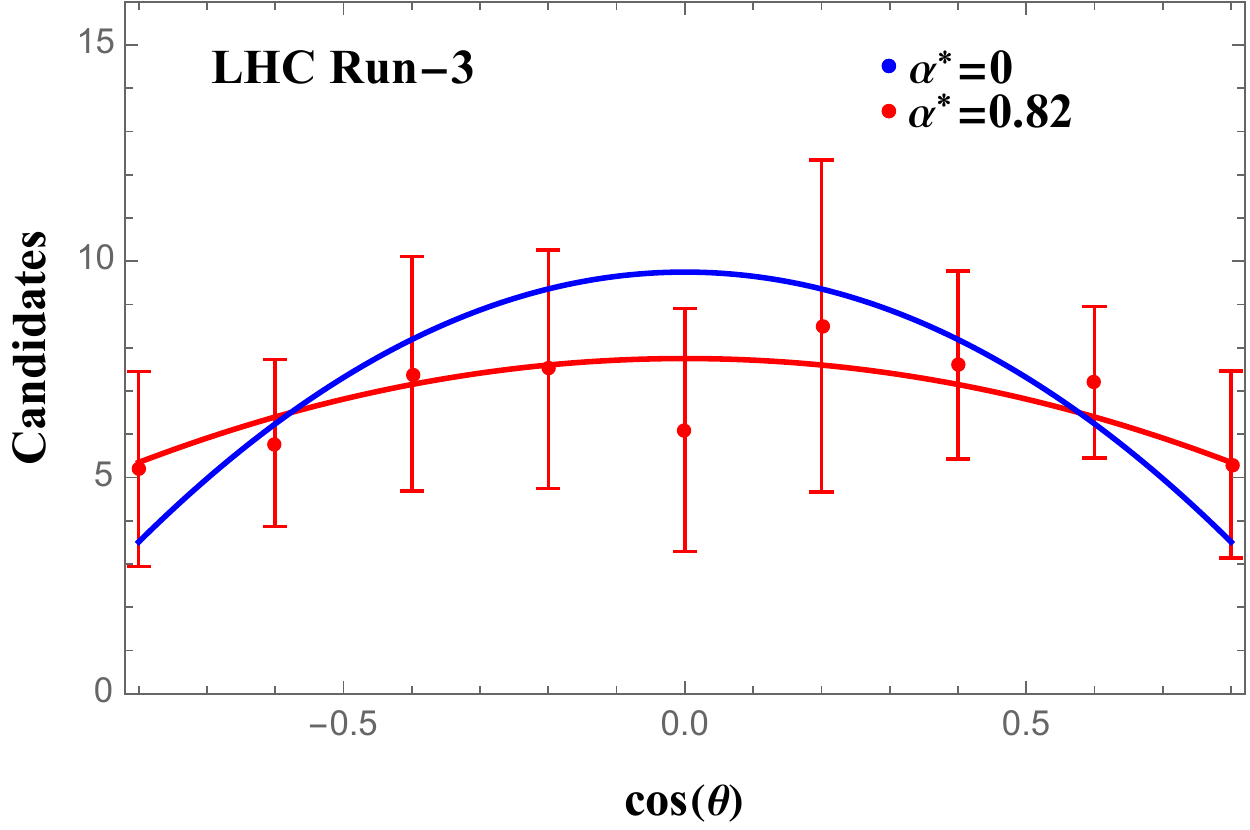}
 	\includegraphics[width=0.45 \textwidth ]{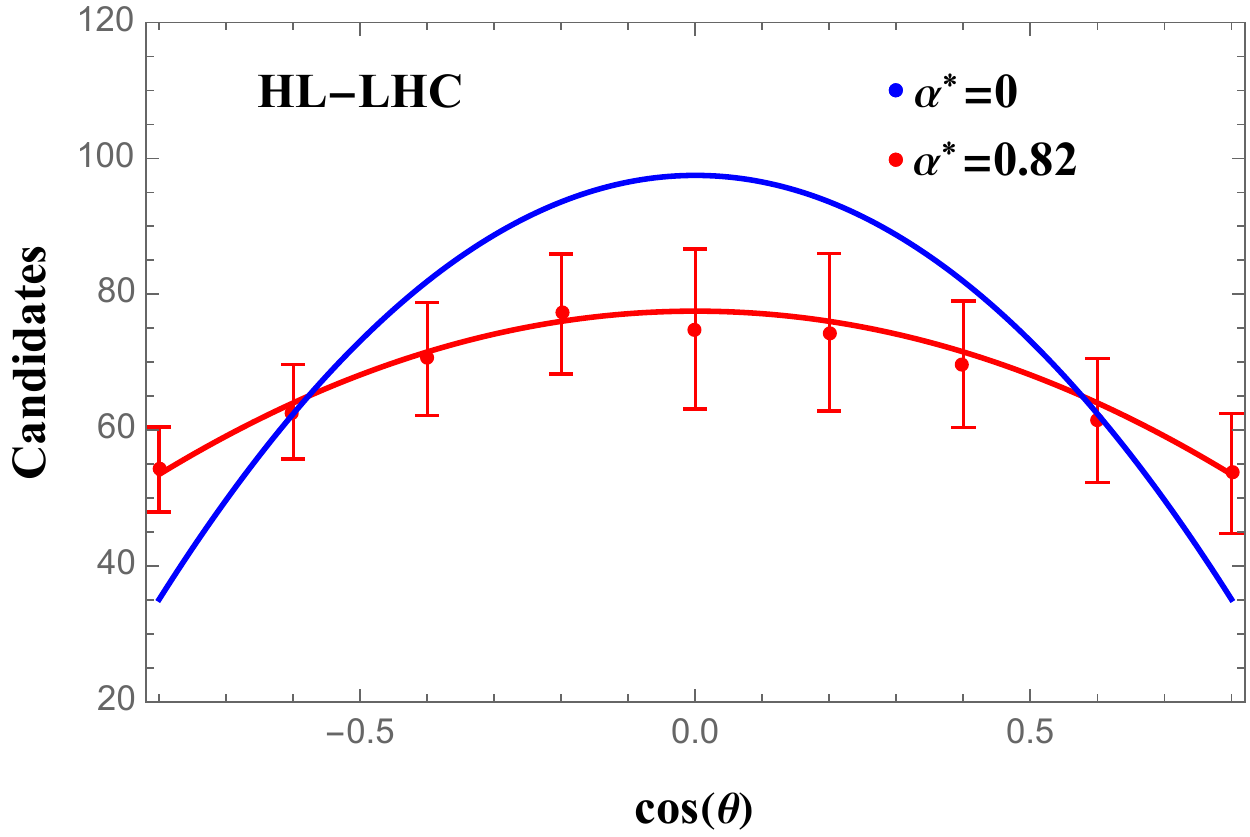}
 	\caption{ 
 		The numbers of the observed events of $B_c^{(\ast)} \to J/\psi(\to l^+l^-) \pi^+$ plotted against $\cos \theta$.
 		The red points with statistical uncertainties are the pseudodata generated by the Monte Carlo method for $\alpha^\ast=0.82$, and the blue and the red lines are drawn with $\alpha^\ast=0$ and $\alpha^\ast=0.82$ in Eq.~\eqref{alphabar}, respectively.  } 
 	\label{angular_distributions}
 \end{figure}

We generate the pseudodata based on the experimental conditions at LHC Run-2, LHC Run-3, and HL-LHC. 
The off-shell contributions from $B_c^+$ are also included with $N_{B_c} = N_{B_c^\ast}$ as discussed in the  previous paragraph.
The numbers of the events are plotted against $\cos \theta$ in Fig.~\ref{angular_distributions}\,,
 and the numerical results of  $\overline{\alpha}$ and $ \overline{ {\cal A} }$ are given in Table~\ref{table2}\,. 
Our analysis show that  there would be a $1.5\sigma$  signal of nonzero   $\overline{{\cal A}}$ at LHC Run-2 , and  a $5\sigma$ signal at HL-LHC, which would be a solid evidence of $B_c^*$. 
 
\begin{table} [h]
	\caption{The $\overline{\alpha}$ and $\overline{{\cal A} }$ fitted from the pseudodata in  FIG.~\ref{angular_distributions} with statistical uncertainties. }
	\label{table2}
	\begin{tabular}[t]{lcccc}
		\hline
		\hline
	&$N_{B_c} = N_{B_c^\ast}$& $\overline{\alpha}$&$\overline{{\cal A} }$\\
	\hline
LHC Run-2&33& $0.48\pm 0.39$ &$  0.15\pm 0.10$\\
LHC Run-3&65 &$0.43 \pm 0.26$ &  $ 0.17 \pm 0.07$ \\
HL-LHC &650 & $ 0.44 \pm 0.10 $ & $  0.19 \pm 0.03$  \\
\hline
\hline
	\end{tabular}
\end{table}

On the other hand, at the forthcoming experiments at FCC-hh~\cite{FCC:2018vvp}, the number of $B_c^\ast $ events are expected to be  $10^{12}$. 
Hence, there would be about $2000$ and $10^5$ $B_c^{\ast +} \to B^+ \phi$ events at HL-LHC and FCC-hh, respectively, which would be sufficient for the experiments to determine the mass. 

\section{Conclusions}\label{conclusion}
Utilizing the conservation laws,
we propose two novel methods for distinguishing $B_c^*$ and $B_c$ in the experiments. 
The calculated branching fractions of $B_c^+ \to J/\psi \pi^+$ and $B_c^{*+} \to B^+ \phi $ are compatible with the literature, indicating that our analysis is reliable. The nonzero polarized fraction of $J/\psi$ from $B_c^{(*)+} \to J/\psi \pi^+$  has been found to be $\alpha = 0 ~(0.82)$. Furthermore, the branching fractions of $B_c^{*+} \to J/\psi \pi^+$ and  $B_c^{*+} \to \phi \pi^+$ have been obtained as  $(2.4\pm 0.5 ) \times 10^{-8}$ and $(7.0\pm 3.0) \times 10^{-9}$, respectively. To calculate the lifetime of $B_c^*$,  we have found that $\Gamma( B_c^{*+} \to B_c^+ \gamma) = (53\pm 3 )$ eV with the homogeneous bag model, consistent with most of  the literature.

We have shown that  $B_c^{*+} \to B^+\phi$ would be promising to be measured at HL-LHC as well as FCC-hh. 
To examine the feasibilities of the measurements, we have  conducted  simulations based on the experimental conditions. 
Remarkably, we have shown that  the helicity analysis on $B_c^{(*)+}\to J/\psi \pi^+$ is ready to be performed  at LHC. 
Thus, we urge the experimentalists to probe the angular distributions of  $B_c^{(*)+}\to J/\psi(\to l^+l^-) \pi^+$ in the region of  $M(J/\psi \pi^+)> 6325$~MeV, which can be served as an evidence of $B_c^{(*)+}$.

\vspace{0.0cm} {\bf Acknowledgments}

The authors would like to acknowledge the helpful discussion with Chao-Qiang Geng, Ying-Rui Hou and Cong-Feng Qiao.

\appendix 
\section*{Appendix : The baryon wave functions} 
Here,
we give the meson wave functions of the homogeneous bag model, which are used in the calculation of the transition matrix elements in the main text. 
In the original version of the bag model, both the asymptotic freedom and the confinement of the QCD are  described by the bag radius, $R$. The quarks are confined in the bag but moving freely within it, satisfying the free Dirac equation
\begin{equation}\label{inside}
	(i\gamma^\mu \partial _\mu - m ) \psi = 0 \,~~~~\text{for}~ r<R\,.
\end{equation}
For low-lying hadrons, we can take the wave functions to be spherical, and   we arrive at
\begin{equation}\label{quark_wave_function}
	\psi(x) _q=\phi_q(\vec{x}) e^{-iE_qt}= N\left(
	\begin{array}{c}
		\omega_{q+} j_0(p_qr) \chi\\
		i\omega_{q-} j_1(p_qr) \hat{r} \cdot \vec{\sigma} \chi\\
	\end{array}
	\right)e^{-iE_qt}\,~~~~\text{for}~ r<R\,,
\end{equation}
where $q$ is the quark flavor, $N$ the normalizing constant, $\chi$ the two component spinor,  $p_q$  the magnitude of the 3-momentum, and $\omega_{q\pm} =\sqrt{1 \pm m_q/E_q}$ with $E_q$ the quark energy. The antiquark wave functions are obtained by taking the charge conjugate.

At the boundary of the bag the current shall vanish, which give us the boundary condition, read as
\begin{equation}\label{boundary}
	\hat{r}\cdot \left( \overline{\psi } ~\vec{\gamma} ~\psi \right)= 0 \,,~~~~\text{at}~ |\vec{x}|=R\,.
\end{equation}
In analogy to the familiar infinite square well, $p_q$ is quantized, satisfying
\begin{equation}\label{momentum_condition}
	\tan (p_qR) = \frac{p_qR}{1-m_qR-E_qR}\,.
\end{equation}
We concern the low-lying hadrons only and therefore take the minimum of $p_q$. At the massless and the heavy quark limits we have 
\begin{equation}
	\lim_{m_qR \to 0} p_qR= 2.0428 \,, ~~~~\lim_{m_qR \to \infty} p_q R=\pi\,,
\end{equation}
respectively.  A meson can be constructed by confining a quark  and  an antiquark to 
a same bag. By considering the bag energy, zero point energy, and the interaction between quarks, the bag model can successfully explain most of the low-lying hadron masses as well as the ratios of the magnetic dipole moments \cite{Zhang:2021yul}.  

However,
despite the success on the hadron masses, the wave functions of the bag model are problematic when it comes to decays. As the description of a static bag is essentially localized, the hadron wave function cannot be the momentum eigenstates, and thus the transition matrix elements cannot be consistently calculated.  This problem has been resolved with the linear superposition of infinite bags by one of the authors~(Liu), and with it the experimental branching ratios of $\Lambda_b \to \Lambda_c^+ \pi ^+ $ and $\Lambda_b \to  p \pi ^+ $ can be well  explained~\cite{Geng:2020ofy}.

In the homogeneous bag model, the meson wave functions at rest are given as 
\begin{equation}\label{wave_function_inrest}
	\Psi(x_{q_1},x_{q_2}) ={\cal N}\int 
	d^3 \vec{x}
	\phi_{q_1}(\vec{x}_{q_1} - \vec{x})  
	\phi^c_{q_2}(\vec{x}_{q_2} - \vec{x}) e^{-i( E_{q_1}t_{q_1}+ E_{q_2}t_{q_2}) } \,,
\end{equation}
where ${\cal N}$ is the normalizing constant, and $c$ in the superscript denotes the charge conjugate. 
The wave function in Eq.~\eqref{wave_function_inrest} is manifestly invariant under the space translation and therefore describes a meson at rest. The wave functions with nonzero momenta can be easily obtained by Lorentz boost.

By demanding the normalization condition
\begin{equation}\label{pp'}
	\langle p|p'\rangle  = 2 p^0 (2\pi)^3\delta^3(\vec{p}-\vec{p}')\,,
\end{equation}
we find
\begin{equation}
	\frac{1}{{\cal N}^2}  = 2 M \int d^3\vec{x}_\Delta\prod_{i=1,2} d^3 \vec{x}_{q_i}^{\,r} \phi_{q_i}^\dagger\left(\vec{x}^{\,r}_{q_i}+ \frac{1}{2} \vec{x}_\Delta \right)\phi_{q_i}\left(\vec{x}^{\,r}_{q_i}  - \frac{1}{2} \vec{x}_\Delta \right)\,,
\end{equation}
with $p$ and $M$ the hadron momentum and mass, respectively. 

With the wave functions, the meson transition matrix elements can be computed straightforwardly.
For simplicity we take $B_c^{(*)-} \to J/\psi \pi^-$ as an example. The results of 
$B_c^{(*)+} \to J/\psi \pi^+$ can be obtained by taking the CP conjugate as CP is conserved in  the $b\to c $ transition. The transition matrix elements
read as
\begin{eqnarray}\label{vector}
&&	\int \langle J/\psi  |  \overline{c}\gamma^\mu  b(x)  e^{ip_\pi x}|B_c^{(*)-} \rangle d^4 x= {\cal Z} \int d^3\vec{x}_\Delta V^{\mu}(\vec{x}_\Delta)  D_{c}(\vec{x}_\Delta)\,,\nonumber\\
&& \int \langle J/\psi  |  \overline{c}\gamma^\mu\gamma_5  b(x)  e^{ip_\pi x}|B_c^{(*)-} \rangle d^4 x= {\cal Z} \int d^3\vec{x}_\Delta A^{\mu}(\vec{x}_\Delta)  D_{c}(\vec{x}_\Delta)\,,
\end{eqnarray} 
with 
\begin{eqnarray}
	&&{\cal Z} = (2\pi )^4 \delta^4\left(p_{B_c^{(*)} } - p_{J/\psi} -p_\pi\right)    {\cal N}_{B_c^{(*)} } {\cal N}_{J/\psi} \,,\nonumber\\
	&&D_{c}(\vec{x}_{\Delta}) = \sqrt{1-v^2} \int d^3 \vec{x} \phi_{c}^\dagger \left(\vec{x} +\frac{1}{2}\vec{x}_{\Delta}\right) \phi_{c} \left(\vec{x} -\frac{1}{2}\vec{x}_{\Delta}\right)
	e^{-2iE_c  \vec{v}\cdot\vec{x}}\,,\nonumber\\
	&&V^\mu (\vec{ x}_\Delta)=\int  d^3\vec{x} \phi_c^\dagger\left(\vec{x} + \frac{1}{2}\vec{x}_\Delta \right)\gamma^0 \gamma^\mu \phi_b\left(\vec{x} - \frac{1}{2}\vec{x}_\Delta \right) e^{i(M_{J/\psi} + M_{B_c^{(*)} }-E_c-E_b)\vec{ v}\cdot\vec{ x}     }\,,\nonumber\\
	&&A^\mu (\vec{ x}_\Delta)=\int  d^3\vec{x} \phi_c^\dagger\left(\vec{x} + \frac{1}{2}\vec{x}_\Delta \right)\gamma^0 \gamma^\mu \gamma_5\phi_b\left(\vec{x} - \frac{1}{2}\vec{x}_\Delta \right) e^{i(M_{J/\psi} + M_{B_c^{(*)} }-E_c-E_b)\vec{ v}\cdot\vec{ x}     },
\end{eqnarray}
Here, the calculation is taken at the Briet frame where $B_c^-$ and $J/\psi$ have the velocity $-\vec{ v}$ and $\vec{ v}$\,, respectively. 
Although the derivation is quite tedious~(see Ref.~\cite{Geng:2020ofy} for an example), their physical meaning can be easily understood:
\begin{itemize}
	\item  ${\cal Z}$ is the overall normalizing constant along with the momentum conservation.
	\item   $D_c$ is the overlapping coefficient attributed by the spectator quark between  the initial and the final states. Note that their centers of the bags are separated at a distance of $\vec{ x}_\Delta$.
	\item  $V^\mu~(A^\mu)$ is the matrix element of the (axial) vector current at the quark level, where the centers of the bags are separated at a distance of $\vec{ x}_\Delta$.
\end{itemize}
Here,
the exponential in the integrals  would oscillate violently at large velocity, causing a suppression that is a punishment for not being at the same speed.
The matrix elements of $B_c^{*+}\to B^+ \phi$ can be calculated in the same manner.

\end{document}